\documentclass[aps,prb,tightenlines,superscriptaddress,twocolumn,floatfix,longbibliography]{revtex4-1}  %,preprint

\usepackage{graphicx,color}
\usepackage{dcolumn}
\usepackage{booktabs}
\usepackage[colorlinks=true,linkcolor=blue,citecolor=blue,linkcolor=blue]{hyperref}%
\usepackage[normalem]{ulem}

\usepackage{amsmath,amssymb,amsfonts,bm}     %  ,bbm
\usepackage[mathcal]{eucal}
\usepackage{dsfont}

\allowdisplaybreaks[4]

\definecolor{myGreen}{RGB}{46,139,87}

\begin{document}

%-----------------------------------------------------------------
% documentation    title,  authors,   abstract,   pacs
%-----------------------------------------------------------------

\title{Competition-Induced Sign Reversal of Casimir-Lifshitz Torque: An Investigation on Topological Node-Line Semimetal}

\author{Liang Chen}
\email[Corresponding Email:]{slchern@ncepu.edu.cn}
\affiliation{School of Mathematics and Physics, North China Electric Power University, Beijing 102206, China}
\affiliation{Institute of Condensed Matter Physics, North China Electric Power University, Beijing 102206, China}
\affiliation{Hebei Key Laboratory of Physics and Energy Technology, North China Electric Power University, Baoding 071003, China}

\author{Xuan Guo}
\affiliation{School of Mathematics and Physics, North China Electric Power University, Beijing 102206, China}
\affiliation{Institute of Condensed Matter Physics, North China Electric Power University, Beijing 102206, China}

\begin{abstract}
The dispersion of quasiparticles in topological node-line semimetals is significantly different in different directions. In a certain direction, the quasiparticles behave like relativistic particles with constant velocity. In other directions, they act as two-dimensional electron gas. The competition between relativistic and nonrelativistic dispersions can induce a sign reversal of Casimir-Lifshitz torque. Three different approaches can be applied to generate this sign reversal, i.e., tuning the anisotropic parameter or chemical potential in node-line semimetal, changing the distance between this material and substrate birefringence.
Detailed calculations are illustrated for the system with topological node-line semimetal Ca$_3$P$_2$ and liquid crystal material 4-cyano-4-n-pentylcyclohexane-phenyl.
\end{abstract}

\maketitle

\newcommand{\angstrom}{\textup{\AA}}

%-----------------------------------------------------------------
% The body of the paper
%-----------------------------------------------------------------
The Casimir effect \cite{MilonniPW1994,Bordag2009Casimirbook,Klimchitskaya2009RMP} is a pure quantum phenomenon of  vacuum. This effect demonstrates that the zero-point quantum fluctuation of the electromagnetic field can result in a force between two neutral plates on the mesoscopic and nanoscopic scales. The Casimir effect plays a vital role in micro- and nano-electromechanical systems and dominates the fabrication and performance of devices in these systems \cite{ChanHB2001Science,ChanHB2001PRL,BuksE2001PRB,BuksE2001EPL,JavorJ2021MSNE}.

The Casimir-Lifshitz torque (CLT) is another related effect, which considers not only the zero-point energy fluctuation but also the angular momentum of virtual photons. The analytical description of CLT between uniaxially anisotropic half-spaces was established by Barash \cite{Barash1973RQE, Barash1978RQE}. In subsequent investigations, CLT for different birefringent materials has been studied \cite{vanEnk1995PRA,MundayJN2005PRA,ShaoCG2005PRA,MorgadoTA2013OE,Somers2015PRA,GueroutR2015EPL,SomersDAT2017PRL,SomersDAT2017PRA,AntezzaM2020PRL}. Ingenious experiments are proposed to measure CLT, such as utilizing the reorientation of liquid crystal nematics \cite{Somers2015PRA}, using the optically levitated nanorod \cite{XuZ2017PRA}. The precision measurement in the system with liquid crystal 4-yano-4$^{\prime}$-pentylbiphenyl (5CB) and birefringent material demonstrates the existence of CLT \cite{Nature2018Somers}. CLT provides an additional approach for the manipulation of Casimir physics and has substantial potential applications, such as noncontact gears \cite{CaveroPelaez2008PRD,AshourvanA2007PRL}, torsional Casimir actuation \cite{TajikF2017JAP,TajikF2018PRE}, Casimir rotor \cite{MartinezJC2018AIPAdv}, noncontact transfer of angular momentum at nanoscale \cite{Sanders2019CommPhys}, etc. The sign of CLT for a fixed twisting angle may depend on the distance. Recent works give systematic investigations on the sign reversal of CLT between black phosphorus and birefringent materials \cite{Thiyam2018PRL,ThiyamP2019PRB,ThiyamP2022PRB}. However, it is still an open question whether there is a systematic method to find appropriate materials for the observation of CLT sign reversal. In this Letter, we study the CLT between a topological node-line semimetal (TNLSM) and a liquid crystal. We demonstrate that the distinct band structure and dispersion of TNLSM provide a natural mechanism for realizing the CLT sign reversal.

TNLSM \cite{BurkovAA2011PRB} is a special material in which the Fermi surface is a node-line in the three-dimensional Brillouin zone. First principle calculations show that TNLSM can exist in various materials \cite{CarterJM2012PRB,WengHM2015PRB,XieLS2015APLMat,ChanYH2016PRB,YuR2015PRL,KimY2015PRL,LiR2016PRL,WangJT2016PRL,YamakageA2016JPSJ,DuY2017npjQM,XuQ2017PRB,NieSM2019PRB}. Recent experimental measurements show evidence of TNLSM in candidates PbTaSe$_2$ \cite{Bian2016NatComms}, ZrSiSe and ZrSiTe \cite{HuJ2016PRL}, TiB$_2$ \cite{LiuZ2018PRX}, CaAgAs \cite{Takane2018npjQM}, SrAs$_3$ \cite{LiS2018SciBull}, GdSbTe \cite{Hosen2018SciRep}, and Co$_2$MnGa \cite{BelopolskiI2019Science}. Intriguing properties, such as unique Landau energy level \cite{RhimJW2015PRB}, special collective modes \cite{YanZ2016PRB}, long-range Coulomb interactions \cite{HuhY2016PRB}, the drumhead-like surface states \cite{YuR2015PRL,KimY2015PRL}, etc, makes TNLSM attract great attention. Condensed matter materials provide excellent platforms for the experimental observation of novel properties of relativistic particles, e.g., Klein tunneling \cite{YoungAF2009NatPhys,StanderN2009PRL}, Veselago lens \cite{CheianovVV2007Science}, chiral anomaly \cite{ZyuzinAA2012PRB,HuangX2015PRX,ZhangCL2016NatComm}, etc. TNLSM is an even more interesting material where the bulk quasiparticles possess both the properties of relativistic and nonrelativistic particles, i.e., in some directions, the current operator is proportional to the wavevector, in the other direction, the current operator is proportional to a material-dependent \emph{relativistic} velocity. To the best of our knowledge, physical implications which focus on these significant differences have not been reported. The sign reversal of CLT between TNLSM and birefringence is a physical consequence of this property.

\begin{figure}[tb]
	\centering
	\includegraphics[width=\linewidth]{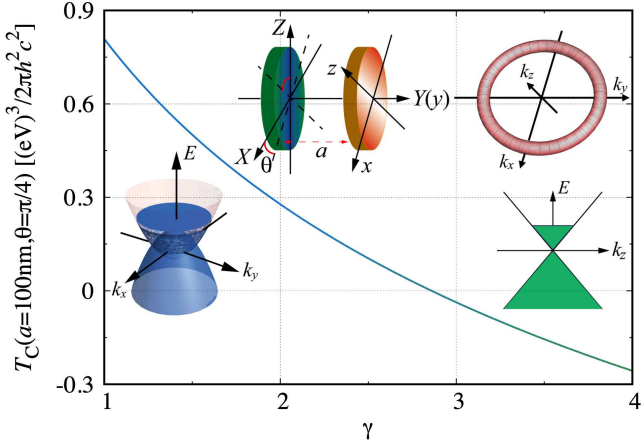}
	\caption{Changing the anisotropic parameter $\gamma$ induced sign reversal of CLT at twisting angle $\theta=\pi/4$ and fixed distance $a=40$ nm. Other parameters for numerical calculation: $\varepsilon_0=0.184$ eV refers to the energy of quasiparticles on the node-line, chemical potential $\mu=0.5\varepsilon_0$, damping parameter $\hbar\eta=\varepsilon_0$. The top left insert shows the geometry of the Casimir interaction between 5PCH and TNLSM. The axes frames $(\hat{X}, \hat{Y}, \hat{Z})$ and $(\hat{x}, \hat{y}, \hat{z})$ are pinned to the principal directions of 5PCH (Green) and TNLSM (Orange), respectively. The top right insert shows the node-line loop structure in the Brillouin zone. The lower inserts show nonrelativistic ($E$ vs $k_x$-$k_y$) and relativistic ($E$ vs $k_z$) dispersions of quasiparticles. These two distinguished dispersions dominate the Casimir torque for anisotropic parameters $\gamma\ll1$ and $\gamma\gg1$, respectively.}
	\label{fig1}
\end{figure}

We consider the CLT between a liquid crystal material 4-cyano-4-n-pentylcyclohexane-phenyl (5PCH) and a TNLSM, as shown in the upper left insert of Fig. \ref{fig1}. The surfaces of 5PCH and TNLSM are located at $y=0$ and $y=a$, respectively. The axes $(\hat{\bm{X}}, \hat{\bm{Y}}, \hat{\bm{Z}})$ and $(\hat{\bm{x}}, \hat{\bm{y}}, \hat{\bm{z}})$ are fixed to the principal directions of 5PCH and TNLSM, respectively. $\theta$ is the twisting angle between $\hat{\bm{X}}$ and $\hat{\bm{x}}$. The node-line loop signature of TNLSM in the wavevector space is illustrated in the upper right insert, where the node-line is located in the $k_z=0$ plane.

The Casimir-Lifshitz torque per unit area is
\begin{equation}
T_C(a,\theta)=-\frac{\partial{E}_C(a,\theta)}{\partial\theta}, \label{eq1}
\end{equation}
%where $a$ and $\theta$ are the distance and twisting angle between the materials separated in the vacuum.
where $E_C(a,\theta)$ is the corresponding Casimir energy density. At finite temperature, it takes the following form according to the Casimir-Lifshitz formula \cite{Somers2015PRA,SomersDAT2017PRA},
\begin{gather}  % (a,\theta)
{E_C}=\frac{k_BT}{4\pi^2}\sum_{n=0}^{\infty}{}^{\prime}\int{\mathrm{d}\bm{p}_{\parallel}}\log\det\left[\mathds{I}-e^{-2\kappa{a}}\mathds{R}_1\mathds{R}_2\right], \label{eq2}
\end{gather}
where $k_B$ is the Boltzmann's constant, $T$ is the temperature, $\bm{p}_{\parallel}$ is the electromagnetic field wavevector parallel to the interface, $\kappa=\sqrt{\bm{p}_{\parallel}^2+\zeta_n^2/c^2}$, $\zeta_n=2n{\pi}k_BT$ is the Matsubara frequency, the prime in the summation means that the term $n=0$ contains a prefactor $1/2$. $\mathds{I}$ is a $2$-by-$2$ identity matrix, $\mathds{R}_1(\bm{p}_{\parallel},i\zeta_n)$ and $\mathds{R}_2(\bm{p}_{\parallel},i\zeta_n)$ are the two Fresnel coefficient matrices on the surfaces.

To obtain a systematic investigation that accommodates the influence of the anisotropic parameter, the chemical potential and the distance, we choose the following SU(2) spin-rotation symmetric two-orbital low energy effective model to describe the specific dispersion of quasiparticles in TNLSM \cite{HuhY2016PRB,BaratiS2017PRB,BurkovAA2018PRB},
\begin{equation}
\hat{H}=\frac{\hbar^2}{2m}(k_x^2+k_y^2-k_0^2)\hat{\tau}_x+\hbar{v_z}k_z\hat{\tau}_y, \label{eq3}
\end{equation}
where $\hbar$ is the reduced Planck's constant, $m$ is the effective mass of the quasiparticles in the $xy$ plane, $k_0$ is the radius of the node-line loop in the Brillouin zone, $v_z$ is the Fermi velocity along the $z$-axis, $\hat{\tau}_x$ and $\hat{\tau}_y$ are the two Pauli matrices acting on the orbital degree of freedom.
In the following context, we reveal sign reversal of CLT at zero temperature. We have calculated CLT at finite temperature by considering both the thermal corrections of permittivity functions and the finite temperature distribution of virtual photons. We find that the finite-temperature corrections can be neglected here.

Utilizing the standard linear response theory, we can derive the analytical expression of permittivity tensor in the imaginary-time formalism. The permittivity tensor of TNLSM is the sum of intraband and interband contributions,
\begin{equation}
\varepsilon_{\alpha\alpha}(i\zeta)=1+\mathcal{K}[\varepsilon_{\alpha\alpha}^{D}(\omega)]+\mathcal{K}[\varepsilon_{\alpha\alpha}^{I}(\omega)],  ~~~ (\alpha=x,y,z),  \label{eq4}  % \mathcalmss
\end{equation}
where $\mathcal{K}[\varepsilon(\omega)]=\frac{2}{\pi}\int_0^{\infty}{\omega\textrm{Im}[\varepsilon(\omega)]}/({\omega^2+\zeta^2})\mathrm{d\omega}$ takes the Kramer-Kronig transformation of the permittivity function, $\varepsilon(\omega)$, expressed in the real frequency representation.
The intraband contribution takes the form of the Drude model, which describes the metallic properties of TNLSM in the infrared region,
\begin{gather}
\mathcal{K}[\varepsilon_{\alpha\alpha}^{D}(\omega)]=\frac{g_{\alpha\alpha}^D}{\zeta(\zeta+\eta)}, \label{eq5}
\end{gather}
where $\eta$ is the damping parameter, $g_{\alpha\alpha}^D$ ($\alpha=x, y, z$) is the Drude weight. $g_{\alpha\alpha}^D$ take the following forms at zero temperature,
\begin{eqnarray}  %. (\mu-\varepsilon_0)
g_{xx}^D&=&\frac{2e^2k_0\mu}{\gamma\hbar^2}\left\{1+\frac{1}{\pi}\left[\frac{1}{6}\sin2\varphi(4-\cos^2\varphi)-\varphi\right]\right\}, \label{eq6} \\
g_{yy}^D&=&g_{xx}^D, \label{eq7} \\
g_{zz}^D&=&\frac{e^2\gamma{k_0}\mu}{\hbar^2}\left[1+\frac{1}{\pi}\left(\frac{1}{2}\sin2\varphi-\varphi\right)\right], \label{eq8}
\end{eqnarray}
$e^2=4\pi/137$ is the fine structure constant, $\gamma=2mv_z/\hbar{k_0}$ describes the spatial anisotropy of the band structure, $\mu$ is the chemical potential, $\varphi=\Theta(\mu-\varepsilon_0)\cos^{-1}(\varepsilon_0/\mu)$, $\varepsilon_0=\hbar^2k_0^2/2m$ defines an energy scale for calculation, $\Theta(\mu-\varepsilon_0)$ is the Heaviside step function. The interband terms take different forms from the standard Ninham-Parsegian oscillator model \cite{ParsegianVAbook}
(the analytical expressions are given in the Supplemental Material).

The competition between relativistic and nonrelativistic dispersions of quasiparticles can be altered by tuning the anisotropic parameter $\gamma$. For both the intraband and interband contributions, we find the following results.
In one strong anisotropy limit $\gamma\rightarrow0$, the nonrelativistic property of TNLSM dominates, both $\mathcal{K}[\varepsilon_{xx}^D(\omega)]$ and $\mathcal{K}[\varepsilon_{xx}^I(\omega)]$ tend to infinity, while $\mathcal{K}[\varepsilon_{zz}^D(\omega)],~\mathcal{K}[\varepsilon_{zz}^I(\omega)]\rightarrow0$. The material tends to the ideal metallic grating limit \cite{MorgadoTA2013OE,GueroutR2015EPL,AntezzaM2020PRL}, i.e., $\varepsilon_{xx}\gg\varepsilon_{zz}$. In the other strong anisotropy limit $\gamma\rightarrow\infty$, we get the opposite result, $\varepsilon_{xx}\ll\varepsilon_{zz}$. Changing the value of $\gamma$ from zero to infinity behaves like a $\pi/2$ rotation of the principal axes of TNLSM. This can drive the CLT to change sign. Figure \ref{fig1} shows CLT at twisting angle $\theta=\pi/4$ and distance $a=100$ nm as a function of the anisotropic parameter $\gamma$ (see the following context and the Supplemental Material for details).  We find that CLT changes sign at $\gamma\approx2$ for the given parameters.

\begin{figure}[tb]
	\centering
	\includegraphics[width=\linewidth]{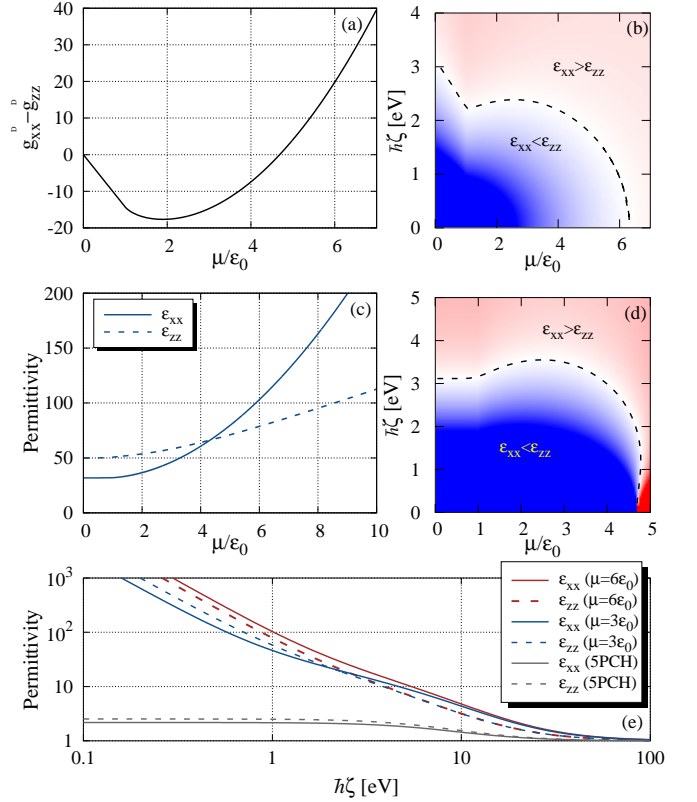}
	\caption{\textbf{Permittivity functions.} (a) The difference between Drude weights, $g_{xx}^{D}-g_{zz}^{D}$, as a function of chemical potential $\mu$. (b) $\mathcal{K}[\varepsilon_{xx}^I(\omega)]-\mathcal{K}[\varepsilon_{zz}^I(\omega)]$ as a function of chemical potential $\mu$ and imaginary frequency $\hbar\zeta$. (c) $\varepsilon_{xx}(i\zeta)$ and $\varepsilon_{zz}(i\zeta)$ given in equation (\ref{eq4}) as a function of chemical potential for imaginary frequency $\hbar\zeta=1$ eV. A cross uccors at $\mu=4.33\varepsilon_0$. (d) Contour plot of $\varepsilon_{xx}(i\zeta)-\varepsilon_{zz}(i\zeta)$ as a function of chemical potential and imaginary frequency. The dashed line shows the boundary where $\varepsilon_{xx}(i\zeta)-\varepsilon_{zz}(i\zeta)=0$. (e) Permittivity of liquid crystal 5PCH and TNLSM ($\mu=3\varepsilon_0$ and $6\varepsilon_0$, respectively) as functions of imaginary frequency. In numerical calculation, the anisotropic parameter $\gamma$ has been set to be 2.8, the energy scale $\varepsilon_0$ has been chosen to be $0.184$ eV. The dielectric functions of 5PCH can be found from Refs. \cite{WuST1993OptEng,KornilovitchPE2012JPCM}. }
	\label{fig2}
\end{figure}

Changing the chemical potential, $\mu$, is another appropriate approach to realize the CLT sign reversal. From equations (\ref{eq6}) and (\ref{eq8}), we find that when $\gamma>\sqrt{2}$ and $\mu<\varepsilon_0$, $g_{xx}^D<g_{zz}^D$. In the other limit $\mu\gg\varepsilon_0$, we get asymptotic expressions,
\begin{gather}
g_{xx}^D=\frac{2e^2k_0\mu}{\gamma\hbar^2}\left[\frac{4}{3\pi}\frac{\mu}{\varepsilon_0}+\frac{1}{2}+\frac{1}{6\pi}\left(\frac{\varepsilon_0}{\mu}\right)^2+...\right], \label{eq9} \\
g_{zz}^D=\frac{e^2\gamma{k_0}\mu}{\hbar^2}\left[\frac{1}{2}+\frac{2}{\pi}\frac{\varepsilon_0}{\mu}-\frac{1}{3\pi}\left(\frac{\varepsilon_0}{\mu}\right)^2+...\right],  \label{eq10}
\end{gather}
such that $g_{zz}^D{\ll}g_{xx}^D$. There is a crossover from $g_{xx}^D<g_{zz}^D$ to $g_{xx}^D>g_{zz}^D$ as $\mu$ increases. Figure \ref{fig2}(a) shows $g_{xx}^D-g_{zz}^D$ as a function of the chemical potential. There is a transition at $\mu/\varepsilon_0=4.68$ for $\gamma=2.8$. We need to emphasize that, although the Drude weights depend on  $k_0$, $\mu$, $\varepsilon_0$, and $\gamma$, the critical point $\mu/\varepsilon_0$ where $g_{xx}^D(\mu/\varepsilon_0)=g_{zz}^D(\mu/\varepsilon_0)$ depends only on the anisotropic parameter $\gamma$. The interband contribution to the permittivity is more complicated. The sign of $\mathcal{K}[\varepsilon_{xx}^I(\omega)]-\mathcal{K}[\varepsilon_{zz}^I(\omega)]$ depends not only on the chemical potential but also on the imaginary frequency $\zeta$. Figure \ref{fig2}(b) shows the numerical result of $\mathcal{K}[\varepsilon_{xx}^I(\omega)]-\mathcal{K}[\varepsilon_{zz}^I(\omega)]$ as a function of $\mu/\varepsilon_0$ and $\hbar\zeta$. The dashed line refers to the boundary where $\mathcal{K}[\varepsilon_{xx}^I(\omega)]=\mathcal{K}[\varepsilon_{zz}^I(\omega)]$. Figures \ref{fig2}(c) and \ref{fig2}(d) show the competition of the total dielectric functions
%(with both contributions from the intraband and interband terms contained)
along different principal axes. In Fig. \ref{fig2}(e), we plot the dielectric functions of 5PCH and TNLSM for typical chemical potentials, $\mu=3 \varepsilon_0$ and $\mu=6 \varepsilon_0$. When the chemical potential is smaller than a threshold (i.e., for $\gamma=2.8$, $\mu=4.78\varepsilon_0$ as shown in Fig. \ref{fig2}(d)), e.g., $\mu=3\varepsilon_0$ as shown by the blue lines in Figure \ref{fig2}(e), there is a reversal of the principal axes at $\hbar\zeta\approx2$ eV,  % 100nm
which is consistent with the boundary as shown in Figure \ref{fig2}(d).  The CLT should reverse at some distance $a_c{\sim}c/\zeta_c$ \cite{Thiyam2018PRL}, where $c$ is the speed of light in the vacuum, and $\zeta_c$ is the critical value of imaginary frequency where $\varepsilon_{xx}(i\zeta_c)-\varepsilon_{zz}(i\zeta_c)=0$. In this case ($\mu=3\varepsilon_0$), we expect $a_c\approx100$ nm.
 When the chemical potential is greater than the threshold, e.g., $\mu=6\varepsilon_0$ as shown by the red lines in Figure \ref{fig2}(e), $\varepsilon_{xx}(i\zeta)-\varepsilon_{zz}(i\zeta)$ has a definite sign, like the conventional birefringent material (the gray lines in Figure \ref{fig2}(e) for 5PCH). The distance-dependent sign reversal of CLT disappears in this case.

On the basis of the above investigations on the permittivity functions, we study the CLT. We derive the analytical expressions of the Fresnel matrices in Eq. (\ref{eq2}) using the standard transfer matrix method \cite{BerremanDW1972JOSA,Gerrard1994TMMbook}.  In Cartesian coordinates,
\begin{gather}
\mathds{R}_1= (\kappa\mathds{U}_1-\mathds{Q}\mathds{V}_1)(\kappa\mathds{U}_1+\mathds{Q}\mathds{V}_1)^{-1},  \label{eq11}\\
\mathds{R}_2= (\kappa\mathds{U}_2+\mathds{Q}\mathds{V}_2)(\kappa\mathds{U}_2-\mathds{Q}\mathds{V}_2)^{-1},   \label{eq12}
\end{gather}
where $\mathds{Q}=\mathds{Q}(\bm{p}_{\parallel}, i\zeta)$ is a 2$\times$2 matrix that determines the propagation of the electromagnetic field in the vacuum, $\mathds{U}_j$ and $\mathds{V}_j$ are the $2$-by-$2$ submatrices of $\mathds{W}_j$ ($j=1,2$) in the following form,
\begin{equation}
\mathds{W}_j=\left( \begin{matrix}
\mathds{U}_j & \mathds{U}_j \\
\mathds{V}_j & -\mathds{V}_j
\end{matrix} \right).  \label{eq13}
\end{equation}
The analytical expressions of $\mathds{W}_1$ and $\mathds{W}_2$ consist of the four eigenmodes of Maxwell's equations in  5PCH and TNLSM, respectively (see the Supplemental Material for detailed expressions).

\begin{figure}[tb]
	\centering
	\includegraphics[width=\linewidth]{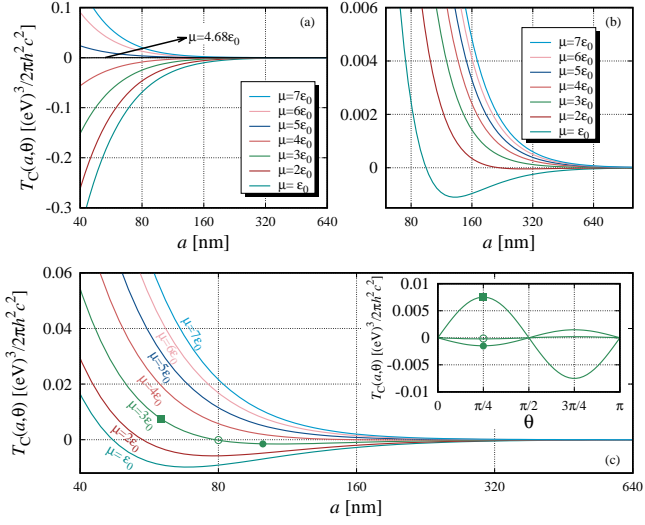}
	\caption{\textbf{Casimir-Lifshiz torque between TNLSM and 5PCH.} (a) CLT as a function of distance where the permittivity functions contain only the intraband contribution. The black line shows the CLT for the isotropic situation where $g_{xx}^{D}(i\zeta)=g_{zz}^{D}(i\zeta)$. (b) CLT as a function of distance where the permittivity functions contain only the interband contribution. (c) CLT as a function of distance for different chemical potentials where the permittivity functions contain both the Drude and interband contributions. The insert shows the CLT vs. twisting angle for $\mu=3\varepsilon_0$ and typical distances, $a=60$ nm ({\tiny\color{myGreen}{$\blacksquare$}}), $80$ nm ({\color{myGreen}{$\circ$}}), and $100$ nm ({\color{myGreen}{$\bullet$}}). }
	\label{fig3}
\end{figure}

As shown in Fig. \ref{fig2}, the Drude and interband terms have significantly different contributions to the permittivity functions. Their contributions to the CLT should be different. We consider the CLT for these terms respectively. Figs. \ref{fig3}(a) and \ref{fig3}(b) show the maximal Casimir torque at $\theta=\pi/4$ as a function of distance with different chemical potentials for the Drude and interband terms, respectively. The black line for $\mu=4.68\varepsilon_0$ shows the boundary where $T_C(a,\theta)$ is exactly zero. For a lower chemical potential ($\mu<4.68\varepsilon_0$), the relativistic dispersion dominates, making the CLT negative.  When $\mu>4.68\varepsilon_0$, the non-relativistic dispersion dominates, $T_C(a,\pi/4)>0$.  The interband contributions are significantly different. In the short distance region ($a<80$ nm), $T_C(a,\pi/4)>0$ for any  chemical potential. This is consistent with the permittivity functions shown in Figure \ref{fig2}(b), where the quadratic dispersion dominates over the linear dispersion in the large imaginary frequency region. In the large distance region, $T_C(a,\pi/4)$ is negative for chemical potential $\mu<5\varepsilon_0$. The sign of CLT reverses at some critical distance, for example, when $\mu=\varepsilon_0$, the critical distance is about $85$ nm. In Figure \ref{fig3}(c), we plot CLT as a function of distance with both the Drude and interband contributions contained. The critical distance is 47 nm for $\mu=\varepsilon_0$ and 79 nm for $\mu=3\varepsilon_0$. For fixed chemical potential $\mu=3\varepsilon_0$, the insert shows the twisting angle dependence of CLT for different distance $a=60$ nm, 80 nm and 100 nm, respectively.

\begin{figure}[tb]
	\centering
	\includegraphics[width=\linewidth]{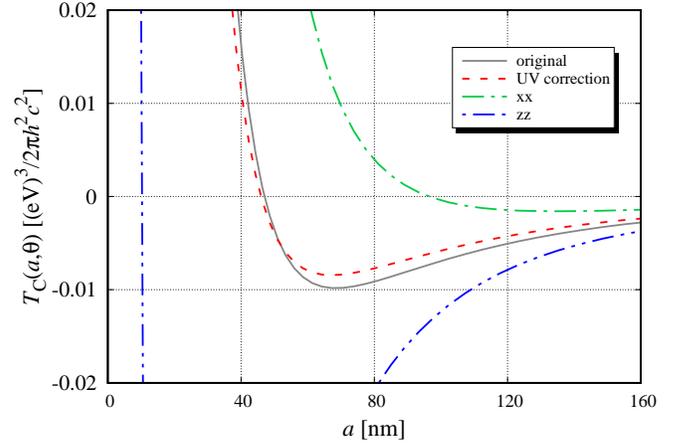}
	\caption{\textbf{UV optical oscillator corrections to the CLT}. The red dashed line shows the CLT as a function of distance when the permittivity function, Eq. (\ref{fig4}), has an additional isotropic UV optical oscillator correction, $g_{UV}/(1+\zeta^2/\omega_{UV}^2)$, where the oscillator frequency $\omega_{UV}$ and oscillating strength $g_{UV}$ have been set to be $3$ eV and $8$, respectively. For comparison, the gray (solid) line, green (dotted) line, and the blue (double-dotted) line show the CLT for the following configurations: without UV optical oscillator, with UV optical oscillator for only the $\varepsilon_{xx}$ component, with UV optical oscillator for only the $\varepsilon_{zz}$ component. $\mu=\varepsilon_0$ in TNLSM is set in calculation. }
	\label{fig4}
\end{figure}

\begin{figure}[tb]
	\centering
	\includegraphics[width=\linewidth]{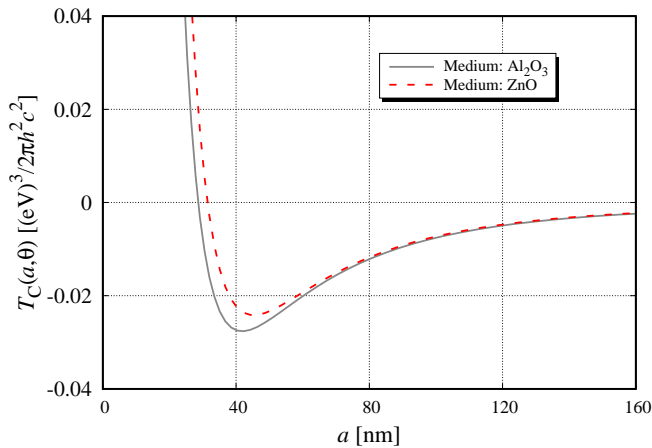}
	\caption{CLT as a function of separation distance between 5PCH and TNLSM with intervening homogeneous layers. $\mu=\varepsilon_0$ in TNLSM is set in calculation. Dielectric functions of Al$_2$O$_3$ and ZnO can be found from Ref. \cite{BergstromL1997ACIS}.}
	\label{fig5}
\end{figure}

The parameters used in this work, that is, the radius of node-line loop $k_0$, the anisotropic parameter $\gamma$, and the effective mass $m$, are fitted for the candidate TNLSM Ca$_3$P$_2$ \cite{ChanYH2016PRB,BaratiS2017PRB}. In this work, we consider only the low energy effective model of TNLSM, the permittivity functions with ultraviolet (UV) optical oscillator corrections may change the Casimir interaction in the short-distance region. Numerical evaluation of CLT by considering the UV optical oscillator shows that the competition-induced sign reversal is valid in a wide parameter regime. In Fig. \ref{fig4}, we calculate CLT by considering UV optical oscillators for three different cases, i.e., both the in-plane and out-of-plane components have corrections (the red dash line), only the in-plane component has correction (the green dotted dash line),  and only the out-of-plane component has correction (the blue double-dotted dash line). We find that the sign reversal appears at about 50, 90, and 10 nm for these cases, respectively.  Furthermore, the critical distances in the interval $45\sim80$ nm (shown in Fig. \ref{fig3}) can be further reduced by intervening homogeneous dielectrics between the slabs \cite{SomersDAT2017PRL,ThiyamP2022PRB}. Figure \ref{fig5} shows CLT at $\theta=\pi/4$ as a function of distance for $\mu=\varepsilon_0$ with Al$_2$O$_3$ and ZnO inserted in between 5PCH and TNLSM. The sign reversal appears at 28 nm and 31 nm for Al$_2$O$_3$ and ZnO intervened respectively.

The sign reversal of CLT is induced by the unique dispersion of TNLSM. The liquid crystal 5PCH can be replaced by other birefringent materials. We calculate the CLT between TNLSM and other birefringent materials, i.e., liquid crystal 5CB, inorganic materials BaTiO$_3$, calcite, and quartz (see Supplemental Material for details). We find that changing the anisotropic parameter, chemical potential, and distance can reproduce the sign reversal of CLT. However, the critical point where sign reversal occurs does depend on the perticular choice of the other birefringent materials in front of TNLSM.

In summary, we study the CLT between TNLSM and conventional birefringent materials. The distinct band structure and dispersion of TNLSM provide a precise mechanism for the sign reversal of CLT, i.e., the relative strengths of polarizabilities along different principal axes can be changed by the competition of relativistic and nonrelativistic properties of quasiparticles in TNLSM. Experimentally, this sign reversal can be manifested by changing the anisotropic parameter, chemical potential, and the distance between TNLSM and birefringent material. Quantitative calculations show that, in the system consists of TNLSM  Ca$_3$P$_2$ and liquid crystal 5PCH, the sign reversal of CLT appears at a separation distance which is very close to the experimental accessible regime.

%\section{Acknowledgement}
The authors are grateful for the financial support from the National Natural Science Foundation of China (Grant No. 12174101) and the Fundamental Research Funds for the Central Universities (Grant No. 2022MS051).

% Specify following sections are appendices. Use \appendix* if there
% only one appendix.

%-----------------------------------------------------------------
% Sec**: References
%-----------------------------------------------------------------
%\nocite{*}

%-----------------------------------------------------------------
% The bibliography
%-----------------------------------------------------------------

%\bibliographystyle{apsrev4-2}
%\bibliography{ref}

%apsrev4-2.bst 2019-01-14 (MD) hand-edited version of apsrev4-1.bst
%Control: key (0)
%Control: author (8) initials jnrlst
%Control: editor formatted (1) identically to author
%Control: production of article title (0) allowed
%Control: page (0) single
%Control: year (1) truncated
%Control: production of eprint (0) enabled
%

\end{document}